\documentclass{article}
\usepackage{graphicx} 
\usepackage{url}
\usepackage[T1]{fontenc}

\usepackage{xcolor}
\title{Causality, determinism and predictability: Probing nature through  classical \\ and quantum eyes}
\author{Mundur V N Murthy\footnote{The Institute of Mathematical Sciences (Retired), Chennai 600113}, 
\ and  A R Usha Devi\footnote{Department of Physics, Bangalore University (Retired), Bengaluru 560056}} 
\date{\today}

\begin{document}
\maketitle
\begin{abstract}
In this article, mainly focussed on students and interested readers, we discuss how the understanding  of causality, determinism and predictability is transformed as we move from the classical world view to the quantum description. We first explain  how determinism and predictability operate in classical dynamics and outline the notion of causality. A paradigm shift occurs within the framework of quantum mechanics where we have to revisit, and even give up,  the notions of determinism, predictability and maybe even strict causality. An engrossing and fascinating debate ensues, after the formulation of quantum mechanics among the giants of physics like Einstein, Bohr, Heisenberg, Pauli, and many others. While some questions have been settled in the last century in favour of quantum world view, there are still many open questions, most important of which is the role of gravity in shaping events within the quantum framework. 
\end{abstract}
\section{Introduction}

 In the beginning of his course of lectures on Quantum Mechanics, R.~Shankar begins by saying ``Now I have got bad news and good news. Bad news is that the subject is hard to follow intuitively and the good news is that nobody can follow it intuitively. Richard Feynman, one of the big figures in physics, used to say that no one understands Quantum Mechanics...''\footnote{ You can listen to him in this lecture: \url{https://www.youtube.com/watch?v=uK2eFv7ne_Q&list=PLozLiCENL19jyrZvpUvG3W79bHKUOY79m}}  Rhetorical flourish apart, there is a ring of truth in the above quotation. We do understand the highly successful mathematical framework of quantum mechanics. We can derive its predictions and test them experimentally with remarkable accuracy. Yet, at a deeper level, many remain dissatisfied, particularly when the thinking is framed by our experience of the world around us fits the classical description very well. However, quantum mechanics forces us to reconsider concepts that appear naturally intertwined in classical physics. In this essay, we try to disentangle what is knowable and what is not within the framework of quantum mechanics from a pedagogic point of view.

As we discuss in the next section, causality, determinism and predictability are closely associated in our everyday understanding of the physical world, but they are not synonymous.  \emph{Causality} refers to the relation between causes and their effects. \emph{Determinism} concerns whether the state of a physical system, together with the laws governing its evolution, uniquely fixes its subsequent state. \emph{Predictability}, in contrast, depends on  whether the future evolution or outcomes of a system can be reliably inferred from the information available to us. Although these notions are closely related, they are not equivalent. Classical mechanics strengthened their association by providing a framework in which a specified physical state evolves according to definite dynamical laws. The development of nonlinear dynamics and chaos, however, showed that deterministic laws need not imply practical predictability.

Quantum mechanics introduces a further distinction: The quantum state evolves according to precise dynamical laws, yet the outcomes of individual measurements are generally described only probabilistically. These developments raise questions that go beyond the mathematical formulation of physical laws. What do we mean by a cause when the outcome of an individual event cannot be uniquely predicted? Does determinism necessarily imply causality? How should we understand the emergence of the familiar classical world from an underlying quantum description? We will examine some of these questions and their consequences.

\section{Classical world view}

The classical world view grew out of human experience of the macroscopic world. Observing positions and velocities of objects  
and the cause-effect connections, resulted in the formulation of the laws of classical dynamics. The success came with every prediction verified through observations and experiments. From the emergence of Newtonian mechanics in the late seventeenth century until the beginning of the twentieth century, our world view was largely framed by these successes.

The mathematical framework of Classical Dynamics as formulated by Newton defined this world view in which, under appropriate conditions, the state of a system determines its subsequent evolution uniquely. The precise statement of Newton's second law of motion that the ``rate of change of momentum is equal to the net force applied on a body in motion" is famously written as $$\vec{F}=m\,\vec{a}=m\,\frac{d^2\vec{r}}{dt^2},$$ where $m$ is the mass and $\vec{a}$ is the acceleration when the mass is assumed to be a constant.  In this form, the equation of motion is the second derivative of the position vector in time and therefore is immune to the arrow of time. That is,  the equations may  be used to predict the outcome of a present event far into the future or far into the past.  

The Laplacian notion of certainty emerged out of this formulation.   Laplace wrote~\cite{Laplace}:
\begin{quote} ``We ought then to consider the present state of the universe as the effect of its previous state and the cause of that which is to follow.... nothing would be uncertain, and the future like the past, would be open to its eyes." \end{quote}
   It is a strong statement of the material world, in which determinism links the present state of the universe to its past as an effect and to its future as a cause.
Laplace posits that, in theory, we should be able to \emph{predict} the future and reconstruct the past with complete certainty, embodying the principle of causal determinism. Often dubbed as the ``Laplace's demon", Laplace envisions a system that encodes the initial position and momentum of every atom in the universe through which one can compute the whole history of the universe past, present, and future.  Nothing is left to chance! 

This view of certainty breaks down when the limits of measurement and the error propagation is taken into account. Few points to note in this context:
\begin{itemize}
    \item  If the body's initial position and momentum can be stated with infinite precision then the evolution of the state of the system can be uniquely traced in the phase space\footnote{The state of a system at a given time in classical mechanics is defined  by its position and momentum. This is denoted by a point in the state space or more conventionally called the phase space. As the time changes the point traces a curve in the phase space and the evolution of the system is thus represented by a unique continuous curve called the phase curve. The space of all such possible states of the system, phase points,  is the phase space.}.  
    \item Even in the absence of infinite precision, if the equations of motion are linear then the errors in the input values of initial state propagate at the same level or may even go down.  Laplace viewed the world through this  `linear' lens. He assumed that a small error in specifying the initial position would lead to only a small error in the predicted outcome. The system was therefore regarded as well-behaved, stable, and predictable. This  world view of predictability is well realised in linear and/or integrable systems, where the evolution is generally stable against small uncertainties in the initial state. 
    \item However, this  deterministic world view  breaks down when confronting a system whose equations of motion are not only non-linear but also non-integrable. The complete breakdown of uniqueness of evolution and stability of a classical system occurs when confronted with singular situations as it occurs in some billiard problems leading to chaos\footnote{There is a class of problems dealing with scattering of billiard balls in an enclosure. The presence of sharp edges, singular scatters,  causes the breakdown of prediction.}. 
\end{itemize}
The breakdown of predictability was pointed out first by Poincaré and we will discuss this next. 

 Poincaré was among the first to recognise that the Laplacian world view had conflated determinism with predictability. Consider a 
 classical dynamical system whose initial state is fixed. If the equations of motion satisfy the appropriate conditions for existence and uniqueness of solutions, then the specified initial state determines a unique trajectory in the phase space. In this mathematical sense, evolution remains deterministic. Poincaré's analysis of the three-body problem~\cite{Poincare}, however, revealed that deterministic classical dynamics can exhibit extreme sensitivity to initial conditions. Two trajectories arising from arbitrarily close, but distinct, initial states may separate rapidly with time. Thus, even though each initial state determines a unique trajectory, finite uncertainty in the specification of the initial state can make long-term prediction practically impossible. This distinction between the state that the system has and what can be predicted about that state is fundamental: Deterministic evolution does not necessarily imply practical predictability. This phenomenon is now commonly associated with the \emph{butterfly effect}\,\footnote{Refers to the phenomenon in which  a small disturbance caused by the flapping of a wing of a butterfly could result, in principle, in cyclones far away.}. To Poincaré, the universe was a nonlinear web in which even perfectly deterministic laws could lead to an unpredictable future, challenging the possibility of absolute certainty of Laplace. He made the distinction between what the state of the system is and what is knowable about it. This distinction can be expressed in a straightforward manner: In a deterministic classical theory, an exact initial state determines a unique trajectory. However, an actual physical state can only be specified with finite accuracy. In a chaotic system, arbitrarily small differences in the initial state can grow rapidly, so that there is, in general,  a loss of long-term predictability.  Chaos does not destroy determinism: It separates the question of whether the evolution is uniquely determined from that of  whether that evolution can be predicted in practice~\cite{MMU}.

A possible bridge between the two situations is seen in some systems that exhibit chaotic behaviour. There are systems which are called strange attractors, for example, the Lorenz system.  This is deterministic, yet its trajectories can exhibit extreme sensitivity to initial conditions. 
 However, the long term pattern or geometry limits the trajectories to lie in a confined region of the phase space, thus illustrating how deterministic evolution can coexist with limited long-term predictability. More generally, deterministic systems may possess attractors of very different kinds, ranging from a fixed point, as in a damped harmonic oscillator, to the more intricate strange attractors encountered in chaotic systems.

Thus, while the mathematical formulation of classical dynamics is deterministic, it may not always be predictable.  The equations of motion for many classical systems are invariant under time reversal, and therefore do not, by themselves, account for the observed macroscopic arrow of time. This brings into focus the concept of \emph{causality}. Time-reversal invariance implies that, for such systems, the dynamical laws do not intrinsically distinguish between past and future evolution. However, this by itself does not establish a theory of causation. Causation is ordinarily understood in terms of temporal ordering, in which a preceding condition \emph{causes} the subsequent effect.

Thus in deterministic dynamics, the state of the system at an earlier time uniquely determines its subsequent state, although this deterministic dependence need not, by itself, constitute a theory of causation~\footnote{Mathematical determination works in both temporal directions. But causal language is normally asymmetric, i.e., cause $\rightarrow$ effect rather than effect $\rightarrow$ cause.}.  

While deterministic dynamics provides a natural setting for causal descriptions, there are deeper questions about the nature of causality. The human brain is wired to analyse observed events in a strict causal sense already. Does this mean that causality is part of a framework, much like space-time, in which we try to understand the measurements? Much like the internal grammar in languages, \emph{is causality intrinsic to our way of understanding the universe?} The answer, according to  philosopher Immanuel Kant, is that causality is an \emph{a priori} category of the understanding through which we organise and interpret our experience of events, rather than something established by empirical measurements alone. Thus, causality is somehow built into the way the human brain makes sense of events. It is a lens through which events are interpreted. This is a much debated point, but the whole edifice of causality, determinism and predictability is going to change dramatically  when our discussion  moves to quantum dynamics.

\section{Quantum world view}

Quantum mechanics provides the mathematical framework for describing the dynamics of physical systems at the microscopic level and for predicting the outcomes of measurements.  A successful theory should predict the result of ``measurements". Quantum mechanics  is spectacularly successful in this sense. It is probably the most successful theory measured against known facts. Quantum physics is therefore the fundamental framework of all of physics while classical physics is a reliable approximation, which is more intuitive, at scales that we deal with in every day life. Logically it may even appear that we should therefore start with the foundations of quantum physics as the basis for all the phenomena.

The mathematical framework of quantum mechanics is simple and precise. A state of a system\footnote{In non-relativistic quantum mechanics, the state function is the representation of the quantum state of a system in the configuration (position) space, often referred to as the ``wave function'' in most books. We avoid this terminology here since it can be misleadingly interpreted as representing a physical wave, it does not.} is represented by  the symbol $|\psi(t)\rangle$, whose time evolution is uniquely determined by the Schr\"odinger equation
$$ i\hbar\frac{d}{dt}|\psi(t)\rangle=\hat H|\psi(t)\rangle,$$
If the state is known at some initial time $t_0$, its subsequent state at a later time $t$ is given by
$$|\psi(t)\rangle= e^{-i (t-t_0)\hat H/\hbar}|\psi(t_0)\rangle.$$
Thus, the unitary dynamics of an isolated quantum system is  deterministic: A specified initial quantum state determines a unique quantum state at every later time.  The unitary evolution is also stable with respect to small errors in the initial state. In particular, unitary evolution preserves the distance between quantum states\footnote{A fundamental requirement in quantum mechanics is that the inner product between the states under time evolution is preserved. This ensures probability conservation and is achieved when the states evolve under unitary transformations:  $|\psi(t)\rangle=U(t-t_0)\,|\psi(t_0)\rangle$, where $UU^\dagger$=I. This condition is sufficient to derive the Schr\"odinger equation. The preservation of distance between quantum states under unitary evolution is analogous to a corresponding property of Hamiltonian classical dynamics:  Phase-space probability distributions are transported by the same volume-preserving flow, and their  overlap is consequently conserved. Thus, although individual classical trajectories may separate rapidly in a chaotic system, the Hamiltonian evolution of probability distributions preserves the information contained in their initial phase space distributions.}, so small differences in the initial states do not grow with time. This is in contrast to the sensitive dependence on initial conditions encountered in chaotic classical systems discussed in the previous section.

This observation is important in the present context. The state of a particle in classical dynamics is defined by its position and momentum at an instant. The evolution of this state is completely determined by the classical laws.  Its predictability or otherwise depends on the error in specifying the initial state and whether the evolution is linear or non-linear.  The introduction of quantum mechanics does not, by itself, destroy determinism at the level of dynamical evolution of the state function. The essential difference from classical mechanics arises when the quantum state is connected to the outcomes of measurements. The quantum state does not in general determine a unique outcome of an individual measurement. Instead, it determines the probabilities of the possible outcomes. This establishes a distinction between the deterministic evolution of the state and the probabilistic nature of individual measurement outcomes. We now examine how the quantum state is related to the probabilities of these outcomes.

\subsection{Quantum state and measurement outcomes}

There is an important difference between the quantum and the classical descriptions when it comes to measurements. The quantum state, represented in general by a complex state function, is not itself a directly measurable physical quantity. In classical mechanics, the dynamical state is specified in terms of physical variables such as position and momentum, whereas in
quantum mechanics the state is connected to physical observations through measurements of observables.

A measurable physical quantity is represented by a Hermitian operator $\hat O$. Its possible measurement outcomes are the eigenvalues $o_n$ satisfying
\begin{equation}
\hat O|o_n\rangle=o_n|o_n\rangle,
\label{eq:observable}
\end{equation}
where $n$ can take many values depending on the system, discrete or sometimes even continuous.  
If the system is prepared in the state,
\begin{equation}
|\psi\rangle=\sum_n c_n|o_n\rangle
\label{eq:state_cn}
\end{equation}
then a measurement of $\hat O$ yields the outcome $o_n$ with probability
\begin{equation}
P(o_n)=|c_n|^2.
\label{eq:measurement_prob}
\end{equation}
The expectation value 
$$\langle\hat O\rangle=\langle\psi|\hat O|\psi\rangle=\sum_{n} P(o_n)\ o_n$$
is the statistical average of the outcomes obtained from an ensemble of identically prepared systems; it is not, in general, the outcome of an individual measurement.

This leads to a fundamental distinction between prediction in classical and quantum theory. In deterministic classical dynamics, a complete initial state determines the subsequent state. In quantum mechanics, the initial quantum state, together with the choice of measurement, determines the probabilities of the possible measurement outcomes, but, in general, does not determine which particular outcome will occur in an individual measurement. Thus, from a purely mathematical point of view, we understand the basis of quantum mechanics very well — we know the basic axioms, how they are put together, and how to use them to calculate and compare with experiments. But if each individual measurement throws different outcomes, what was the state of the system before the measurement? Is this question even relevant? This generated a deep divide among the pioneers of quantum mechanics. While the advocates of Copenhagen school, Bohr, Heisenberg, Pauli et al, advocated a point of view, it was challenged by Einstein, and even by one of the founders of quantum mechanics, viz. Schr\"odinger. 

In order to understand the different positions taken and issues raised, we need to outline some further features of the standard measurement framework. These are important for the discussion that follows:
\begin{itemize}
\item  In the usual formulation of non-relativistic quantum mechanics, the state can be represented in the position (configuration) basis by a wave function $$\psi(\vec r,t)=\langle\vec r|\psi(t)\rangle$$
and the probability of finding the particle in the volume
$d^3r$ around $\vec r$ is given by 
$$
P(\vec r,t)\,d^3r
=
|\psi(\vec r,t)|^2\,d^3r.
$$
The state is, in general, a superposition of position eigenstates,
corresponding to different possible outcomes of a position
measurement.
\item Similarly, for any observable $\hat O$, the quantum state is, in general, a linear superposition of its eigenstates,
as in Eq.~(\ref{eq:state_cn}), with $c_n=\langle o_n|\psi\rangle$  being the probability amplitudes for the corresponding outcomes
$o_n$.
\item The ``Collapse Postulate'' is a central component of the
standard ``Copenhagen interpretation'' of quantum mechanics: During
the act of measurement of an operator $\hat O$ yielding the outcome
$o_n$, the system abruptly collapses to the corresponding eigenstate
$\vert o_n\rangle$. A subsequent measurement yields the same outcome
with certainty.
\end{itemize}

\bigskip

\fbox{
  \begin{minipage}{\textwidth}
  \section*{The relaity debate}
  \small
 There was a major disagreement among the thinkers of quantum mechanics on what kind of microscopic reality quantum mechanics describes. This is exemplified in the following statements: According to Roger Penrose quantum mechanics is incomplete because it does not explain how the state, which is a superposition of many possibilities, collapses to a single state and remains there subsequently. On the other hand, in the 1994 debate between Roger Penrose and Stephen Hawking on the nature of quantum mechanics~\cite{Hawking}, Hawking took a “positivist” position, in contrast to Penrose's “Platonist” position: Hawking remarked, ``I don’t demand that a theory correspond to reality because I don’t know what it is. Reality is not a quality you can test with litmus paper. All I’m concerned with is that the theory should predict the results of measurements.'' Quantum mechanics does this.  The spirit of `'shut up and calculate'' is then to focus on predicting the results of measurements rather than on questions about an underlying reality which can never be tested through a single measurement.  
 
 It is important to emphasise that although the mathematical framework of quantum mechanics is precise and its unitary time evolution is deterministic. The distinction between deterministic mathematical evolution and the question of whether nature is fundamentally deterministic is  more subtle than the usual historical narrative: Del Santo and Gisin~\cite{Gisin} argue that Erwin Schr\"odinger's views on determinism were considerably more nuanced than is commonly supposed; he had questioned the necessity of determinism even before the advent of quantum mechanics. This adds a historical dimension to the  debate on physicists' commitment to determinism as a fundamental principle of nature, across both classical and quantum world views., the probabilistic interpretation of the quantum state, introduced by Max Born, leaves room for different interpretations of what the quantum state represents physically and strong disagreements as seen in the positions taken by Penrose and Hawking.\end{minipage}
}

\bigskip

\begin{itemize}
\item The full probability distribution, or the weights for each eigenstate in the expansion, can be determined experimentally by
making a large number of identical copies of the system and performing identical measurements on all of them. Each time the collapse occurs abruptly into one of the eigenstates, but once a statistically significant number of measurements are completed, the full probability distribution of the eigenstates in the superposition emerges.
\end{itemize}

A classic example of how these factors influence an outcome is the famous quantum interference experiment. The first clear demonstration of this phenomenon is by Davisson and Germer who sent a beam of electrons on crystalline nickel target. Since the atoms are regularly spaced,  the crystal produces diffraction of the scattered electrons, which are detected at different angles using a rotatable detector.

But the real clincher came with the classic \textit{Hitachi Experiment} by Tonomura and his collaborators~\cite{Tonomura}. The experiment involved sending electrons one by one with a delay, roughly about 10 per second,  towards an electron bi-prism which acts like a double slit. The scattered electrons are captured by a position sensitive detector on which it shows up as a dot. Initially the dots seem to appear in random positions but after a large number of electrons are sent, a clear interference pattern emerges. Note that the electrons are sent individually with delay, so the interference pattern can only emerge if each electron interferes with itself i.e., each state of the electron is a superposition. 

Some thing strange happens when we try to find out through which slit the electron could have passed. This can be done by placing detectors near the slits. However, the moment the electrons are detected at a target location, the interference pattern disappears.  We just get two bright spots corresponding to the electrons passing one by one  through individual slits. The act of detection forces the electrons to \textit{collapse} into one of the eigenstates and the interference pattern disappears.

We are now entering a strange territory where we have to go beyond the mathematical framework into the domain of interpretation, which is contested. The Copenhagen interpretation, associated primarily with Niels Bohr, Werner Heisenberg, and Wolfgang Pauli, emerged from the discussions among these founders of quantum mechanics in the 1920s and 1930s. According to Copenhagen interpretation, the state function is a mere tool to calculate physical quantities and it does not have a physical reality on its own in the sense that it can be measured. The collapse of the state function merely updates the state of the system corresponding to that measurement. Large number of such updates give the complete picture of the system. In fact, the Copenhagen interpretation forces us to assume that there is no objective reality independent of the observer. The reality is in fact defined by the measurement itself. In this view, physical properties acquire definite meaning through measurement, rather than being regarded as possessing pre-existing values independently of measurement\footnote{We digress a little here. In the Buddhist philosophy of Nagarjuna, who lived in the 2nd century CE, the doctrine of \textit{Sunyata}, meaning emptiness, is put forward. Nagarjuna questions whether things possess \textit{svabhava}---an intrinsic and independent nature. The doctrine posits that things are empty of intrinsic characteristics and they do not exist independently of the conditions and relations in which they arise;  Nothing exists in isolation. Things come into existence only in relation to their causes, conditions and relations. Obviously, this does not mean Nagarjuna was expounding quantum mechanics, but it is an interesting parallel 
to  a related debate on \textit{the nature of reality}.}. There are competing interpretations like the one advanced by Penrose, which treats the state function as representing reality, and the many worlds interpretation which does not accept the collapse. We will not go into these interpretations in detail here.

There is now a complete paradigm shift from classical physics. The first victim of this shift is predictability, although the mathematical framework of unitary quantum dynamics remains deterministic.  The evolution of the state is uniquely determined, but the state  is not an observable. Instead it provides probabilistic estimates for the physical quantities. That is, the estimate is only valid when a large number of measurements are made on identical copies of the system. We can not predict the outcome of a single measurement but only provide probable outcome of such a measurement.  This is not simply a practical problem in specifying the state with precision as argued by Poincaré where the Laplacian certainty can only be achieved when there is infinite precision as discussed in the previous section. 
\subsection{Causality becomes a casualty!} 
Even more significant is the collapse of the notion of strict causality as in classical dynamics. As argued in the previous section  deterministic dynamics establishes a unique relation between successive states of a system. The trajectories in the phase space follow  unique paths and never intersect. But determinism by itself does not constitute a theory of causation.  In classical mechanics, this distinction is often obscured because the state of a system at one time uniquely determines its state at another time. 
In the quantum framework, the present follows the past. But one can not predict the precise path in time. Given a state of the system at present, we can only predict the probable states as time evolves. Quantum mechanics  poses a different question about causality: The outcome of an individual measurement cannot, in general, be assigned a unique value on the basis of the quantum state alone. The challenge to causality is therefore not that quantum dynamics ceases to connect earlier and later states, but that the connection between a preparation and an individual observed outcome is no longer deterministic in the classical sense. 

Radioactive decay provides a familiar example. The decay of an unstable nucleus is governed by well-defined quantum dynamics, and the theory predicts the statistical distribution of decay times for an ensemble of identically prepared nuclei. However, the exact time at which a particular nucleus decays is not, within the standard formulation of quantum mechanics, determined by the initial state in the manner of a classical trajectory. The distinction is therefore not between an event with a cause and an event without a cause, but between a physical process governed by definite dynamical laws and the probabilistic prediction of the outcome of an individual event.  

The mathematician and philosopher of science Grete Hermann (a doctoral student of Emmy Noether), argued that the statistical character of quantum mechanics does not, by itself, invalidate the principle of causality~\cite{Hermann1}. The inability to predict the outcome of an individual quantum event with certainty need not imply that causal relations have disappeared; rather, quantum mechanics may require a broader conception of causality in which causal relations need not be expressed as deterministic predictions of future events.

A more striking challenge arises when measurement itself becomes relevant to the subsequent evolution of a quantum system. In classical physics, observation is ordinarily regarded as revealing the state of a system without fundamentally altering its dynamical evolution. Quantum mechanics, however, allows repeated measurements to influence the evolution of the system. This is illustrated by the \emph{quantum Zeno effect} : When an unstable quantum state that would normally evolve into a decayed state is repeatedly measured at sufficiently short time intervals, the measurement projects the system back onto the undecayed state. Repeated projections can therefore inhibit the transition to the decayed state. In the ideal limit of infinitely frequent measurements, the transition can be completely suppressed. This was theoretically analyzed by Misra and Sudarshan~\cite{ECGS} and subsequently demonstrated experimentally~\cite{Itano}. The Quantum Zeno effect presents a more direct challenge to the classical conception of causality. The act of measurement is not merely a passive observation of an independently evolving system; repeated measurement can alter the subsequent evolution itself.
Quantum theory, therefore, challenges the classical identification of causality with a deterministic and continuously evolving trajectory. The question is no longer simply whether every event has a unique predictable outcome, but how causal relations should be understood when physical processes are described quantum mechanically and measurement itself can play an active role in determining subsequent evolution.

The real question then is the following: Does the edifice of quantum mechanics, constructed as above, represent an incomplete picture of reality or it is in fact intrinsic  to the nature of things we observe?  This was the question that animated greatest of minds for many decades, and it still does for many, which we will discuss  next. 

\vskip 0.5cm
\subsection{Einstein versus Bohr: The EPR paradox}

A serious challenge to the Copenhagen interpretation of quantum mechanics came from Einstein and his coworkers. In a 1935 paper published in Physical Review~\cite{EPR}, Einstein, Podolsky and Rosen (EPR) questioned whether the quantum-mechanical description of \emph{physical reality} could be regarded as \emph{complete}. The criterion of objective reality  put forth by EPR is the following: ``\emph{if the value of a physical quantity can be predicted with certainty without in any way disturbing the system, then there exists an element of physical reality corresponding to that quantity.}"\,\footnote{In Ref.~\cite{NM} ``\emph{ element of physical reality}" gets abbreviated, rather delightfully, as ``\emph{epr}"! The capitalized EPR, meanwhile, has long stood for Einstein-Podolsky-Rosen.}  Let us illustrate the question with some examples. We contrast a classical example with the quantum example to drive home the differences.   

Consider the following situation within the framework of classical laws.  Let us  say an asteroid hurtling towards earth disintegrates into two pieces after collision with the upper atmosphere and one of the pieces is detected near earth's surface. We know the original  momentum of the asteroid and now we have also measured the momentum of the piece detected on earth. Conservation law of linear momentum immediately fixes the momentum of the missing piece of the asteroid without any detection.  It is important to note that the momenta of the two pieces of the asteroid are completely determined by the law whether we measure it or not. There is no information exchanged between the two pieces. Thus there exists an objective (pre-existing) reality that is revealed to the observer by such a measurement. 

Once we understand the notion of \emph{objective reality} as a pre-existing fact, we get into conflict with the fundamental interpretation of quantum mechanics.  For example, let us consider the decay of a neutral particle called the pion.
$$\pi^0~\rightarrow~ e^++e^-$$
This is one of the decay modes of the pion, not the dominant one, but it will do for our purposes. The net spin angular momentum of the pion is zero. Therefore to conserve angular momentum the electron and the positron, which are spin-half particles, should have angular momentum projections in opposite directions. That is, if the spin of the electron is measured along any chosen direction and is found to be Up, the positron spin will be found to be Down along the same direction, and vice-versa. So, if an experiment is designed to detect the electron spin state (Up or Down) in a given direction, we immediately know the positron spin state (Down or UP) in the same direction.  The crucial feature of this example is that the electron and positron are not described as two independent quantum systems. Their joint quantum state is an \emph{entangled state}, in which the spin outcomes of the two particles are correlated even when the particles are spatially separated
It is this remarkable feature, viz., \emph{entanglement}\,\footnote{In his  lucid article~\cite{Simon}  Simon puts this rather nicely: \emph{A book written in an entangled state will have to be grasped in its entirety as one unit, and not sentence by sentence or paragraph by paragraph. In this respect an entangled state is like a good piece of art!}}, that makes the outcome of a measurement on one particle perfectly predictable from the corresponding measurement on the other. To see how this quantum correlation leads to the EPR argument, let us now consider what quantum mechanics says about the state of the two particles before and after detection.

Note that before detection, the electron is a superposition of both Up and Down states, that is, it has no well-defined spin state of its own — according to quantum mechanics! During the act of measurement, the electron collapses in-deterministically into one of the two possible spin states, Up or Down. Even more bizarre is the fact that consequent to this measurement on the electron, the positron, no matter how separated it is from the electron, also collapses to Down or Up state respectively. In other words, once the electron spin is measured, the spin of the positron can be predicted with certainty, as a consequence of the correlations in their entangled state, even though the spin of the positron itself has not been measured. According to the EPR criterion of objective reality stated above, this implies that the corresponding spin of the positron must be an element of physical reality, even before it is measured. It was natural for  EPR to assign a definite  value for the positron spin component  as a pre-existing physical reality rather than something created by the measurement. 

But there is a deeper issue here: The direction along which the spin of the electron is measured can be chosen arbitrarily. Therefore, the same argument can be applied to the pre-existing spin component of the positron along any direction. EPR were thus led to the conclusion that the positron must possess definite values of its spin components corresponding to different possible choices of measurement direction at the electron end, even before any measurement is performed. This conclusion is in tension with the quantum-mechanical description, since spin components along different directions are represented by non-commuting operators and satisfy corresponding  Heisenberg uncertainty (also called \emph{indeterminacy}) relations, which  do not allow them to have simultaneous definite values in general. 

We thus arrive at the EPR paradox: Either the quantum-mechanical description is incomplete, or the measurements on the electron somehow determine incompatible properties of the distant positron. These two alternatives are mutually exclusive: If quantum mechanics is considered complete, the positron cannot already possess definite values for its spin components (which is essentially the other alternative assumption). EPR started their discussion with the first option, namely, that quantum mechanics provides a complete description. Yet, by following the logical consequences of their criterion of objective reality (illustrated using the electron-positron example here), they were led to the second alternative: The spin components of the positron must be objective realities. In other words, the very assumption of the completeness of the quantum-mechanical description leads to its opposite!

The EPR argument naturally raises the question: If quantum mechanics is incomplete, what is missing from its description? One possibility is that the quantum state does not contain all the information about the physical system. There may be additional, as yet unknown, variables that determine the outcomes of measurements. Such variables are generally referred to as \emph{hidden variables}. EPR's argument also rests on the assumption that a measurement performed on one particle cannot instantaneously affect the physical reality of a distant particle at a space-like separation. This is the notion of \emph{locality}, which asserts that no physical influence can propagate faster than the speed of light.  The combination of \emph{locality} with the assumption that physical properties can possess measurement-independent objective values is often referred to as \emph{local realism}.

The Copenhagen school,led by Bohr,  rejected the EPR criterion of physical reality and argued that the result of a quantum measurement should not be interpreted to be a property that existed before the measurement. According to Copenhagen interpretation of quantum mechanics there is no experimentally accessible concept of reality beyond what the measurement tells us. The questions raised by EPR therefore had no experimental way of settling the debate and remained largely in the realm of philosophical discussion, exemplified by Einstein's rhetorical question~\cite{Pais}, \textit{Is the moon there when nobody is looking?} This changed dramatically with the work of John Bell in 1964, who showed that the issue could be turned into an experimentally testable question.

\subsection{From EPR to Bell: Hidden variables and the fate of locality}

The EPR argument did not suggest how to construct a more complete description of physical reality. It left open the possibility that the quantum state might be supplemented by additional \emph{hidden variables} that specify the physical state more completely. The idea was to ask whether the probabilistic character of quantum mechanics could arise from an underlying description in which the physical state is more completely specified. The hidden-variable programme can therefore be viewed as an attempt to go beyond the quantum-mechanical description by introducing additional variables that represent physical properties of a system, which are not specified by the quantum state and that may exist independently of measurement.

One of the earliest attempts in this direction was Louis de Broglie’s pilot-wave formulation, developed during 1926–27, well before the EPR argument of 1935. In this theory, the wave function is supplemented by a definite particle configuration. The particle configuration evolves according to a deterministic guidance law, providing a concrete example of how a hidden-variable theory could yield a satisfactory description underlying the quantum dynamics. De Broglie presented this theory at the Fifth Solvay Conference~\cite{deBroglie} in 1927. However, his proposal did not gain acceptance at the time, and he subsequently abandoned the theory. In 1952, David Bohm independently developed a more complete version of the pilot-wave approach, which is now known as Bohmian mechanics (or the de Broglie-Bohm pilot wave theory)~\footnote{The pilot-wave approach has a mathematical antecedent in the hydrodynamic formulation developed by Madelung~\cite{Madelung}. He reformulated Schr\"odinger's equation in terms of a probability density and a velocity field, revealing a hydrodynamic description governed by a continuity equation and a modified Hamilton–Jacobi equation.
Madelung's hydrodynamic analogy  provides a mathematical connection to the pilot-wave approach. However,   by itself, it does not establish the additional ontological claim of de Broglie–Bohm theory that \emph{particles follow deterministic trajectories in conjunction with the evolving Schr\"odinger wave function}.}. Bohm's theory retains the deterministic evolution of the particle configuration, but it is found to be intrinsically non-local\footnote{For a system of several particles, the motion of each particle is guided by the wave function of the entire system. Thus, the motion of one particle depends not only on its own position, but also on the locations of the other particles, even when they are spatially separated.}.

In the meantime, the possibility of hidden-variable theories had been challenged by a mathematical argument due to John von Neumann. In his 1932 book \textit{Mathematical Foundations of Quantum Mechanics}, von Neumann presented what was widely regarded as a proof that no hidden-variable theory could reproduce the predictions of quantum mechanics. However, the assumptions underlying his argument were subsequently questioned.  This issue was recognized relatively early by  Grete Hermann~\cite{Hermann2}, who  gave a detailed criticism of von Neumann's argument.  Hermann's work, largely ignored for many decades,  demonstrated that von Neumann's argument did not rule out the possibility of a deeper causal description of quantum phenomena in terms of hidden variables~\cite{MVNU}. 

Bohm's work~\cite{Bohm} in 1952 showed that a hidden-variable theory reproducing the predictions of quantum mechanics was indeed possible. Bohm constructed an explicit hidden-variable theory in which particles have definite positions and their evolution is deterministic. Once the wavefunction and the initial particle configuration are specified, the subsequent evolution of the configuration is determined by a guidance equation. But its intrinsic non-locality shows that determinism does not by itself restore the classical world view. In other words, a deeper description can retain determinism, but only at the cost of giving up locality!  

This brought the discussion to a more precise question. The issue was no longer simply whether hidden variables, or even a deterministic completion of quantum mechanics, were possible. The deeper question was whether such a completion could also preserve locality. The correlations between measurements on spatially separated entangled systems made the issue of locality precise and testable.  John Bell~\cite{Bell} addressed this conceptual problem in 1964. He showed that any local hidden-variable theory must satisfy certain inequalities, now known as Bell's inequalities,  for such correlations between the outcomes of measurements performed on spatially separated systems. Quantum mechanics, however, predicts violations of these inequalities, and such violations have been confirmed experimentally~\cite{Aspect}.

The historical development therefore leads to an important distinction for our discussion of causality, determinism and predictability: The experimental violation of Bell inequalities does not by itself rule out hidden variables or determinism. Bohmian mechanics provides a counterexample: It is deterministic, but non-local. What Bell's theorem shows is that the correlations predicted by quantum mechanics cannot be reproduced by a local hidden-variable theory satisfying the assumptions of Bell's framework. Thus, the transition from EPR to Bell did not simply establish that ``determinism does not hold.'' Rather, it revealed that the different features of the classical world view can not \emph{all} be retained simultaneously in a theory reproducing the quantum predictions.

\section{The world in quantum eyes!}

Having learnt the basic principles of quantum mechanics, one can solve a variety of problems involving atoms, molecules, photons and other microscopic systems. The mathematical framework is remarkably successful: The Schr\"odinger equation tells us how a quantum state evolves, and the Born rule tells us how to calculate the probabilities of different measurement outcomes. Yet, after learning the formalism, a deeper question often remains unanswered:
\begin{quote}
\emph{What kind of world does quantum mechanics actually describe?}
\end{quote}
Quantum mechanics challenges some of the intuitions that we naturally carry from our experience of the classical world. It presents us with a world of indeterminate individual events, yet with non-local correlations between outcomes of  spatially separated systems.  In particular, the idea that physical properties have definite values before measurement, together with the assumption that \emph{spatially separated systems can be described independently of one another}, cannot be maintained in a straightforward way as suggested by classical physics. 

\bigskip 

\fbox{
  \begin{minipage}{\textwidth}
  \section*{Historical Notes}
\small
The debate over causality, determinism and hidden variables resurfaced in the mid-1940s~\cite{DelSanto}. British mathematician and mathematical physicist Edmund Taylor Whittaker delivered a lecture at the Physical Society in London during 1943~\cite{Whittaker1943}, in which he challenged the classical doctrine of determinism.

Around the same time, Satyendra Nath Bose gave his presidential address to the Indian Science Congress in Delhi during early 1944 on ``The Classical Determinism and the Quantum Theory''~\cite{Bose1944}. Reflecting on the conceptual shift in the scientific worldview, Bose remarked: ``Fifty years ago the belief in causality and determination was absolute. To-day physicists have gained knowledge but lost their faith.''

The addresses by Whittaker and Bose renewed attention to fundamental questions concerning causality and determinism, in view of von Neumann's no-go argument against deterministic hidden-variable theories. The discussion soon developed into a high-profile exchange in the pages of \emph{Nature} during 1944, beginning with British mathematician Piaggio's ``Collapse of Determinism''~\cite{Piaggio1944}. His article prompted a response from experimental physicist G. D. West, who questioned both the claim that causality had been disproved and the reliance on von Neumann's mathematical argument~\cite{West1944}. Piaggio replied in October 1944~\cite{PiaggioReply1944}. The discussion continued in November, when W. W. Barkas and E. T. Whittaker exchanged views on whether apparent randomness necessarily implied indeterminism~\cite{Barkas1944,Whittaker1944}. Further correspondence appeared under the heading ``Causality or Indeterminism?'' in January 1945. In March 1945, Piaggio returned to the subject with a more comprehensive account of the arguments raised in the correspondence~\cite{Piaggio1945}. Reviewing the evidence and the objections to von Neumann's argument, he brought the series to a close without claiming that the question had been settled. The exchange nevertheless made clear that the relationship between causality, determinism and the foundations of quantum mechanics remained an open question.
  \end{minipage}}

\bigskip

The EPR--Bell discussion also teaches us that determinism and locality are logically distinct. A deeper theory may retain determinism while abandoning locality, as illustrated by Bohmian mechanics. Conversely, the experimental violation of Bell inequalities rules out a local hidden-variable description reproducing the quantum predictions. Thus, Bell's result  places a fundamental constraint on how a deterministic underlying description, if one is sought, can be constructed.

The quantum world, therefore, cannot be understood simply by replacing classical equations with quantum equations, while retaining all our classical intuitions. Quantum mechanics forces us to reconsider what we mean by causality, determinism and predictability. The Schr\"odinger equation gives a deterministic evolution of the quantum state, whereas the Born rule gives only probabilities for the outcomes of measurements. The correlations predicted by quantum mechanics cannot be reproduced by a local hidden-variable theory satisfying the assumptions of Bell's framework. Thus, quantum mechanics does not fit naturally into the classical picture in which physical properties are definite before measurement, evolve according to deterministic laws, and can be understood through \emph{local causes}. The precise interpretation of the quantum state and the origin of measurement probabilities, however, remain deeper questions about the meaning of the theory.

At the same time, the classical world is not simply absent from quantum theory. Macroscopic systems interact continuously with their environment, and these interactions can suppress quantum interference between different components of a superposition. This process, known as \emph{decoherence}, can cause certain states to behave effectively as classical states and provides an important mechanism for understanding the emergence of classical behaviour from underlying quantum dynamics~\cite{Zurek}. 

Decoherence, however, does not by itself explain why a particular outcome is obtained in an individual measurement.  The measurement problem therefore remains an important conceptual question: \emph{How does the quantum description of a system and its interaction with a measuring apparatus account for the definite outcomes we observe?} When does a physical quantity acquire a determinate value in a measurement or when do inherent potentialities become actual?
Different interpretations of quantum mechanics address this question in different ways. The many-worlds or relative-state interpretation, originating in Everett's work, retains unitary evolution without wavefunction collapse and interprets the different components of the quantum state as corresponding to different branches of the physical description~\cite{Everett}. Other approaches modify or supplement the standard framework in different ways. Objective-collapse theories~\cite{Bassi}, for example, introduce stochastic physical mechanisms that modify the standard unitary evolution and produce definite outcomes. Superdeterministic approaches~\cite{Hoss}, on the other hand, re-examine the assumptions underlying Bell's hidden-variable framework. In particular, they question the assumption of \emph{measurement independence}, namely that the distribution of the hidden variables is statistically independent of the choices of measurement settings. These approaches illustrate that the conceptual questions raised by quantum mechanics do not have a single  universally accepted resolution.

\section*{Summary}

The conceptual evolution of the notions of causality, determinism, and predictability has undergone significant changes right from their origins in human experience to their modern interpretation in physics. The starting point is the belief that an external physical world exists independently of the observer and that its structure and behaviour can be understood through observation, experimentation, and mathematical description. The remarkable success of classical mechanics, together with later developments in electromagnetism and relativity, strengthened the view that nature is governed by well-defined laws capable of explaining physical phenomena. However, the study of nonlinear systems and chaos revealed that deterministic laws do not always guarantee predictability, thereby revealing distinctions among concepts that were earlier regarded as closely connected.

\bigskip

The foundation of quantum mechanics followed a profound crisis in our
understanding of the physical world. Several phenomena encountered in the
microscopic domain defied the notions of reality and physical behaviour that
had been formed from our experience of the macroscopic world. Their successful
description by quantum mechanics not only challenged the classical worldview
but also raised the more fundamental question: ``What constitutes physical
reality?" The most significant aspect in this context was the concern
raised by Einstein and his collaborators regarding the completeness of quantum
mechanics in the light of their conception of physical reality. The subsequent
discussion of quantum entanglement, Bell's analysis, and the experimental
verification of Bell-type inequalities transformed the debate, which was tending towards a philosophical debate,  into an
experimentally accessible question about local realism and the nature of
physical reality. 
These developments challenged classical assumptions about local realism, and led to the recognition that the correlations predicted by quantum mechanics cannot be reproduced by a local hidden-variable theory. So where does that leave us now?

The mathematical foundation of quantum mechanics stands on a firm footing and is spectacularly successful in predicting experimental outcomes. The Copenhagen interpretation, that  the state function predicts probabilities which are determined after statistically significant number of measurements in each of which the system collapses to one of the predicted states, is accepted as a pragmatic proposition and as such taught in all the introductory courses. This is not, however,  universally accepted as discussed above. 

But the real ``elephant in the room" is the conflict between quantum mechanics and gravity. Quantum mechanics (or more precisely, Quantum Field Theory (QFT)) forms the framework in which we describe  interactions  between fundamental constituents of matter. While the standard model of particle physics based on quantum field theory  is highly successful, what it omits is also extremely important, namely gravity. Here we encounter a serious problem of fundamental incompatibility between quantum theory and  gravitation. 

Gravity is intimately connected with the structure of space-time itself, whereas quantum mechanics is ordinarily formulated within a given space-time framework. Within the framework of quantum mechanics, the background space-time is fixed and is flat. This is in contradiction with Einstein's theory of gravitation. In general theory of relativity formulated by Einstein the space-time dynamically evolves  depending on the {\it location} and distribution of mass and energy.  Again this contradicts the non-locality concept in quantum mechanics.    Therefore, a consistent theory combining quantum mechanics with gravity remains an open problem at present. Thus, the questions concerning the emergence of classical space-time and macroscopic structures cannot be regarded as completely settled within our present theoretical framework. Quantum mechanics remains incomplete until this foundational question is answered!

\section*{Acknowledgements}
We thank K.~S.~Mallesh for his thoughtful questions, discussions and corrections that helped shape this article from its inception. We are grateful to Rajaram Nityananda for his insightful comments and stimulating discussions on several conceptual aspects of this work. We wish to acknowledge C.~S.~Sundar for his valuable comments on the manuscript and suggestions for enhancing its clarity. We also thank B.~N.~Karthik, Radhika Vatsan, M.~S.~R.~Kumar, and Sudha for reading the manuscript and offering useful suggestions.

\end{document}